\documentclass[prl,twocolumn,english]{revtex4-1}
\usepackage{amsmath,amssymb,mathrsfs}
\usepackage{natbib}
\usepackage{subfigure}
\usepackage{tabularx}
\usepackage{epsfig}
\usepackage{longtable}
\usepackage{amsfonts}
\usepackage{rotating}
\usepackage{subfigure}
\usepackage{amsmath}
\usepackage{hyperref}
\usepackage{babel}
\usepackage{comment}

\usepackage{color}

\def\nn{\nonumber}

\newcommand{\bea}{\begin{eqnarray}}
\newcommand{\eea}{\end{eqnarray}}
\newcommand{\la}{\label}
\newcommand{\be}{\begin{equation}}
\newcommand{\ee}{\end{equation}}

\newcommand{\sgn}{\,\mbox{sgn}}



\begin{document}

\title{Many body localization and quantum non-ergodicity in a model with a single-particle mobility edge} 
 \author{Xiaopeng Li}
 \author{Sriram Ganeshan}
 \author{J. H. Pixley}
\author{S. Das Sarma}
\affiliation{Condensed Matter Theory Center and Joint Quantum Institute, Department of Physics, University of Maryland, College Park, MD 20742, USA}

\date{\today}

\begin{abstract}
 
We investigate many-body localization in the presence of a single-particle mobility edge.  By considering an interacting deterministic model with an incommensurate potential in one dimension we find that the single-particle mobility edge in the noninteracting system leads to a many-body mobility edge in the corresponding interacting system for certain parameter regimes.  Using exact diagonalization, we probe the mobility edge via energy resolved entanglement entropy (EE) and study the energy resolved applicability (or failure) of the eigenstate thermalization hypothesis (ETH). Our numerical results indicate that the transition separating area and volume law scaling of the EE does not coincide with the non-thermal to thermal transition. Consequently, there exists an {\it extended non-ergodic phase} for an intermediate energy window where the many-body eigenstates violate the ETH while manifesting volume law EE scaling. We also establish that the model possesses an infinite temperature many-body localization transition despite the existence of a single-particle mobility edge. We  propose a practical scheme to test our predictions in atomic optical lattice experiments which can directly probe the effects of the mobility edge. 
\end{abstract}
\maketitle

Thermalization, a commonplace phenomenon in various physical settings, can naturally fail in isolated disordered quantum interacting systems, making standard concepts of quantum statistical mechanics invalid. The fundamental theoretical underpinning of thermalization in quantum systems has been postulated in the form of the eigenstate thermalization hypothesis (ETH)~\cite{1991_Deutsch_ETH_PRA,1994_Srednicki_ETH_PRE}. Recently, it has been shown using perturbative arguments that the presence of interaction and disorder in a closed quantum system could lead to many-body localization (MBL)~\cite{Basko06} with such an interacting quantum MBL state being non-thermal. 
 
 A hallmark of MBL is its violation of the ETH~\cite{1994_Srednicki_ETH_PRE}, where a local subsystem fails to thermalize with its environment~\cite{2013_Bauer_JSM}. 
 MBL has now been established non-perturbatively in lattice models with finite energy density, where numerical evidence points towards the existence of  MBL all the way to infinite temperature~\cite{oganesyan2007,pal2010}. Further numerical work~\cite{moore2012,vadim,bardarson2014} and a rigorous mathematical proof~\cite{imbrie2014many} for the existence of the MBL phase have mounted compelling evidence for the existence of  such a `finite-temperature' MBL phase which eventually gives way to an extended phase at strong enough interaction. 
 Although much of the MBL work has focused on the interacting one dimensional (1d) fermionic Anderson model with random disorder~\cite{AndersonLocalization} (and closely related spin models), it turns out that MBL also exists without any disorder~\cite{vadim,altshuler,2015_Schiulaz_PRB} for the Aubry-Andre-Azbel-Harper (AAAH) model~\cite{AA, azbel, harper55}, which is a non-random 1d  model with a quasiperiodic 
 onsite potential.  
 We emphasize that neither 1d Anderson model nor AAAH model manifests a single-particle mobility edge (SPME).

In the absence of a SPME, interactions act on the Fock space of Slater determinants of either completely localized or delocalized single-particle eigenstates. Therefore, introducing a SPME allows one to study how localized and delocalized eigenstates will interact, thus introducing qualitatively new physics. There are several deterministic 1d incommensurate models with 
SPMEs in the literature~\cite{griniasty1988localization, sankarprl88, thouless1988localization, sankarprb90, biddleprl10, biddleprb11, sriramgaa}, 
which can be adapted for studying MBL in the presence of  a SPME.

We consider a recent generalization~\cite{sriramgaa}  of the 1d AAAH model with an analytical expression for the SPME, 
which enables us to study the interplay of many-body effects and the SPME in a controlled fashion. Since the MBL phase is a property of all eigenstates, the presence of a mobility edge adds a  new dimension to the problem as both localized and delocalized single-particle orbitals are now present in the problem.  Using exact diagonalization we find for certain parameter regimes of the model:
(1) The existence of a many-body mobility edge ($E_L$) characterized by the area to volume law scaling of entanglement entropy (EE). (2) A distinct energy scale ($E_T$) that separates a thermal (i.e. ergodic) and non-thermal region in energy, which is established by directly considering an ETH violation based on the criterion in Ref.~\cite{1991_Deutsch_ETH_PRA,1994_Srednicki_ETH_PRE,2008_Rigol_thermalization_Nature}. (3) Our results suggest  $E_T \neq E_L$ and consequently the existence of a non-ergodic regime with volume law EE scaling 
between $E_T$ and $E_L$.
All three of our findings are completely novel differing drastically
from previous studies suggesting
 a sharp many-body mobility edge~\cite{bardarson2014,2015_Luitz_MBLedge_PRB,2015_Mondragon_arXiv}. 
 To guide future experiments that could probe our predicted 
 mobility-edge physics, 
we present a realistic scheme with a straightforward modification to the existing experimental setup~\cite{blochmbl,roati08, modugno}.

\begin{figure*}[htp!] 
\includegraphics[angle=0,width=.8\linewidth]{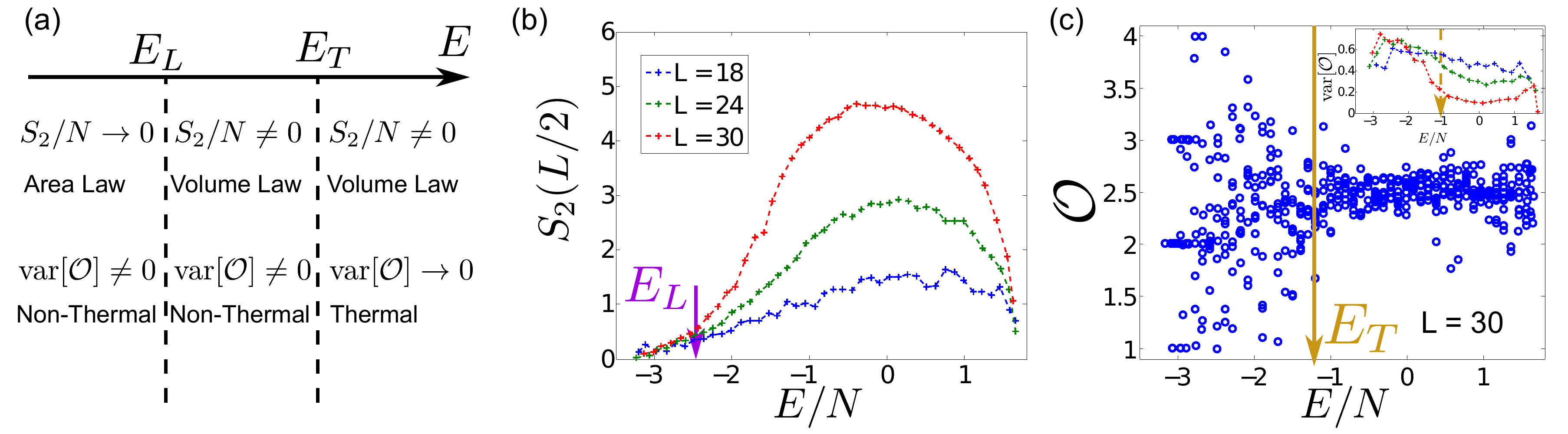}
 \caption{ Energy dependent ergodicity for interacting fermions. (a), A scenario for many-body mobility edge physics as we have found for the GAA model. (b) The bipartite R\'enyi entropy. The eigenstates above $E_L$ have extensive EE, and are thus extended; whereas the states below $E_L$ exhibit EE of area law scaling and are localized. (c) The energy dependence of an observable ${\cal O} (E) $ (Eq.~\eqref{eq:OE}) with $L  = 30$. The inset shows the fluctuation ${\rm var} [{\cal O}]$ (see main text) for different system sizes, using the same plot scheme as in (b). 
The fluctuation of ${\cal O}$ for states $E>E_T$ is small and gets significantly larger for states $E< E_T$. 
 The system is expected to be thermal (non-thermal) above (below) $E_T$. 
(The slight increase of ${\rm var}[{\cal O}]$ near $E/N \approx 1.2$ is an artifact from the small number of states close to the spectra-edge.) 
Our numerical results suggest $E_L <E_T$. In this plot, the filling is fixed at $1/6$, and we use $\lambda/t =0.3$, $V/t =1$, and $\alpha = -0.80$, 
and we average over 
$\phi$ 
for better statistics~\cite{citesupplement}. In the calculation for system size $L = 30$, we use inverse Lanczos and target  $500$ interior eigenstates. }
\label{fig:Fig2} 
\end{figure*}

The model we consider is a generalized Aubry-Andre (GAA) model~\cite{sriramgaa}, 
$H = H_0 + H_{\rm int},$
\begin{eqnarray} 
 \textstyle H_0 & = & \textstyle -t \sum_{j=1}^{L} (c_j ^\dag c_{j+1} + H.c.) +2\lambda\frac{ \cos ( 2\pi q j + \phi) } { 1- \alpha \cos(2\pi q j + \phi) } n_j  \,\,\,\,
 \nonumber    \\
 \textstyle H_{\rm int} &=&  \textstyle V \sum_j n_j n_{j+1}, 
\label{eq:Ham} 
\end{eqnarray}
where $c_j $ is a fermionic annihilation operator, $n_j=c_j^\dag c_j$, and the tunneling $t$ is the energy unit throughout.
We focus only on the fermionic case here. 
We consider $\alpha \in (-1, 1)$, 
with the AAAH model corresponding to  
$\alpha=0$.  
In the non-interacting limit, the GAA model with 
an irrational wavenumber $q$ (we fix $q=2/(1+\sqrt{5})$, 
with no loss of generality), 
has a SPME~\cite{sriramgaa} at $\alpha \epsilon =2 \sgn(\lambda) (|t|-|\lambda|)$~\cite{citesupplement}.  
In this model, the particle number $\sum_j {n_j} = N$ is conserved.

\paragraph*{Interaction effects on localization and thermalization in the presence of SPME.---}
 We study interaction effects on localization and thermalization for the model Hamiltonian $H$ using exact diagonalization. 
In the AAAH model ($\alpha=0$), the non-interacting many-body wavefunction is a Slater determinant of all localized or all extended single-particle orbitals. 
This results in the interacting AAAH model having all many-body states either localized and non-thermal or extended and thermal~\cite{vadim}. 

However, for the non-interacting GAA model (with $\alpha \neq 0$), there are more possibilities originating from the SPME, where the Slater determinant can be composed of both localized and extended single-particle orbitals. Adding interactions to such a system may result in richer many-body states where the localization and thermalization properties may be qualitatively different from the $\alpha=0$ case.  To this end, we employ separate diagnostics to study localization and  
ergodicity without making the common assumption 
that thermalization and delocalization must necessarily be intrinsically connected in an interacting system.
To investigate the localization properties, we cut the lattice at site $l$, which divides it into two subsystems $A$ and $B$, we then calculate the energy resolved R\'enyi entropy $S_2 (l) = -\log ( \mathrm{Tr} \rho_A^2 )$ of  $A$ involving lattice sites $1$, $2$, $\ldots$, $l$, whose reduced density matrix is obtained by tracing out region $B$ at the other sites ($l+1$, $l+2$,$\ldots$, $L$)~\cite{1961_Renyi}.   
The EE scaling reliably tracks the localization transition, where localized and delocalized many-body states are quantified by the area law ($S_2 \sim L^{d-1}$) and volume law scaling ($S_2 \sim L^d$) respectively~\cite{2013_Bauer_JSM, vadim}. 
To understand the thermalization features we calculate the observable ${\cal O} (E)$,
\be 
\textstyle {\cal O} (E) = \sum_{j=1}^{L/2} \langle \Psi_{E} |  n_j  |\Psi_{E} \rangle, 
\label{eq:OE}
\ee 
with $|\Psi_E \rangle$ a many-body eigenstate. The large fluctuation in ${\cal O} (E) $ among eigenstates that are nearby in energy is a signature for the violation of the ETH~\cite{2008_Rigol_thermalization_Nature}.  

We begin by focusing on the MBL transition as a function of energy for fixed model parameters.
In Fig.~\ref{fig:Fig2} we show the energy resolved $S_2(l=L/2)$ for various system sizes.  We find eigenstates with an energy below a certain value $E_L$ are localized with an EE that obeys area law scaling ($S_2(L/2) \sim L^0$)~\cite{2013_Bauer_JSM}, whereas the eigenstates with an energy above $E_L$ exhibit a volume law scaling of the EE ($S_2(L/2) \sim L$) and are thus extended. We define $E_L$ where $S_2(L/2)$ splays out in system size as shown in Fig.~\ref{fig:Fig2}b.  Thus $E_L$ defines the many-body mobility edge, which separates states with an area law scaling from extended states with volume law scaling.
Although the existence of the many-body mobility edge $E_L$ in our model is already a significant result, below we discuss the key issue of whether $E_L$ also defines the ergodic properties of the interacting system.

We now come to the thermalization properties, which are captured by the energy resolved observable ${\cal O} (E)$ 
 as shown in Fig.~\ref{fig:Fig2}c. The fluctuations of ${\cal O}(E)$ within a narrow energy window are quantified by their variance, denoted as  $\mathrm{var}[{\cal O}]$~\cite{citesupplement}.  
 As shown in Fig.~\ref{fig:Fig2}c there is a clear energy threshold $E_T$, that separates two qualitatively different regimes.  
In the energy window ($E>E_T$), the fluctuations of ${\cal O}$ among nearby eigenstates are small, 
for which the ETH is satisfied and the eigenstates are thermal~\cite{2008_Rigol_thermalization_Nature}. 
 The spread of the observable broadens out for energies $E< E_T$, where the fluctuations are significantly larger, leading to a violation of ETH~\cite{2008_Rigol_thermalization_Nature}.
We emphasize here that the thermal to non-thermal transition  is unique to interacting systems, and is absolutely absent without interactions~\cite{2008_Rigol_thermalization_Nature,2008_Rigol_PRA}.

 Our numerical results suggest that the MBL and thermal transitions in energy do not coincide, $E_L \neq E_T$ (see Fig.~\ref{fig:Fig2}). In particular, our numerics imply that $E_T > E_L$, and as a result there exists an energy window ($E_L < E <E_T$) where the many-body states are non-ergodic (violate the ETH) but remain extended (volume law scaling of EE). Therefore, we conclude that in this interacting many-body system with SPME two 
 critical energy scales $E_L$ and $E_T$ exist, which is 
 qualitatively different 
 from the scenario of a sharp many-body mobility edge~\cite{2015_Mondragon_arXiv}. We emphasize that our results are completely distinct from the model in the absence of a SPME ($\alpha=0$), which has no many-body mobility edge and all of the eigenstates are either thermal and delocalized or non-thermal and localized~\cite{citesupplement}. Thus our numerical results point to the existence of a non-ergodic extended regime defined as $E_L < E <E_T$. The possible existence of a non-ergodic extended (i.e. metallic) phase in the vicinity of the MBL transition has been speculated in the literature 
 with no concrete examples~\cite{altshuler1997quasiparticle, 2014_Grover_MBL_arXiv, pino2015non}. 

\begin{figure}[htp!] 
\includegraphics[angle=0,width=\linewidth]{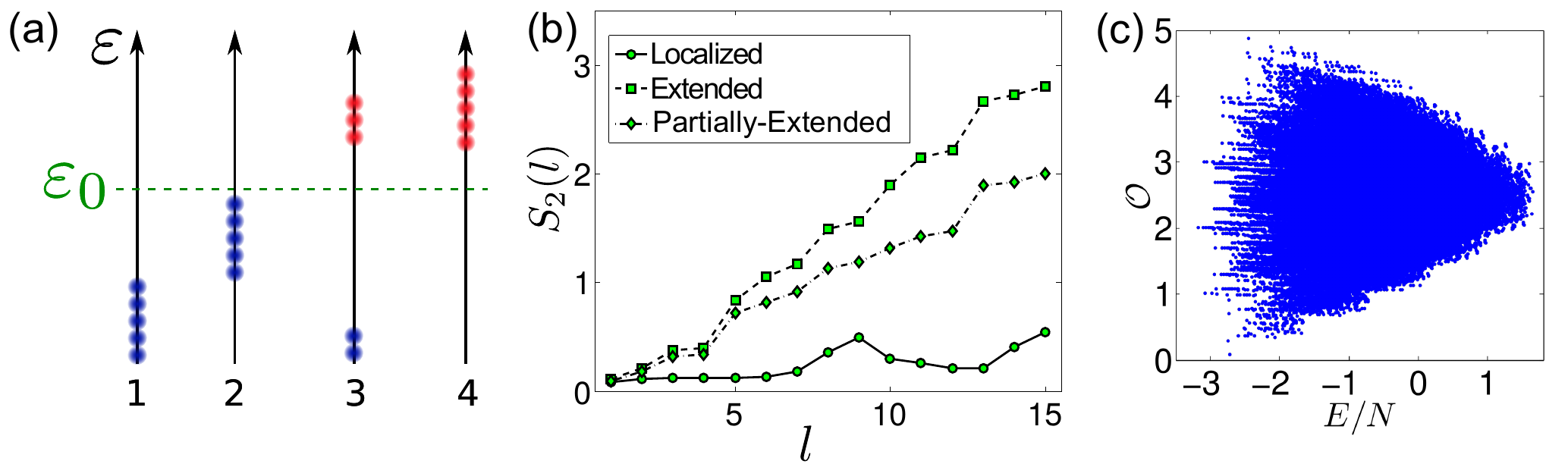}
 \caption{ Localization and non-thermalization of non-interacting fermions with a SPME. Here we simulate $5$ particles in $30$ sites. 
 We  take the GAA model with $\lambda/t = 0.3$, $\alpha = -0.8$, 
 with 
 mobility edge 
 $\varepsilon_0/t = -1.75$. (a) illustrates different possibilities of many-body states. As shown the partially-extended state as marked by $3$ could have approximately the same energy as the localized one marked by $2$. (b) shows the entanglement scaling for the three types of states, localized, extended and partially-extended. The partially-extended states exhibit extensive EE, similar to the extended ones. In (c) we show the energy dependence of ${\cal O}$ (Eq.~\eqref{eq:OE}).
}
\label{fig:Fig1} 
\end{figure}

\paragraph*{Non-interacting many-body states.---}
In the following, we develop a physical intuition for the observed many-body mobility edge. 
The non-interacting many-body states are trivially non-ergodic and violate ETH~\cite{nonthermalnote}, as shown in Fig.~\ref{fig:Fig1}(c)  for $V=0$ where the energy dependence of $\mathcal{O}(E)$
manifests large fluctuations among eigenstates that are nearby in energy~\cite{2008_Rigol_thermalization_Nature}.  
We will discuss the non-interacting limit 
to gain insight into
the emergence of the non-ergodic extended phase in the interacting case.

Without interactions, the many-body eigenstate of $N$ fermions 
is a product state of $N$ single-particle orbitals.  
In the presence of SPME, there are three different ways of constructing many-body states (Fig.~\ref{fig:Fig1}(a)): (i) all particles put in localized single-particle orbitals, (ii) all particles in extended orbitals, and (iii) some particles in localized and others in extended orbitals, which respectively give localized, extended and partially-extended many-body states. 
The partially-extended  states have extensive  EE  (Fig.~\ref{fig:Fig1}(b)). 
(We mention that the partially-extended states would appear localized~\cite{citesupplement} from the perspective of the normalized participation ratio~\cite{vadim}.) Such partially-extended many-body states lead to important consequences. Consider a model with single-particle energies $\epsilon_1 < \epsilon_2 < \ldots < \epsilon_L$ having a SPME $\epsilon_{m^\star}$, 
such that the states  with $\epsilon_{m \le m^\star} $ ($\epsilon_{m > m^\star}$) are localized (extended). 
The lowest energy for a fermionic many-body partially-extended state 
is 
$E_A  =  \epsilon_{m^\star +1} + \sum_{m = 1} ^{N-1}  \epsilon_m$. 
The highest energy of a localized state is $E _B  = \sum_{m = m^\star -N+1} ^{ m^\star} \epsilon_m$. 
For a general Hamiltonian, $E_B>E_A$ is the most typical scenario~\cite{citationnote}. 
The many-body states with energy  $<$($>$) $E_A$ are completely localized (extended, partially-extended or localized); 
the states above $E_B$ are extended or partially-extended.  
In the energy regime $E_B > E > E_A$, however, localized and partially-extended states are coexisting 
by virtue of the SPME in the spectrum which enables the existence of this mixed intermediate energy regime.  

Putting the non-interacting and interacting results all together, a physical picture naturally emerges. Interaction effects on the extended many-body states lead to thermal behavior, whereas for localized states, interactions make them non-thermal. The most interesting case is the coexistence of  partially-extended and localized states, where interactions could stabilize a non-thermal extended phase. 
Such a physical scenario is possible and seems to be consistent with our numerics.  

\paragraph*{The MBL phase in the GAA model.---}  

 Past studies have established the existence of a MBL phase (with all eigenstates localized) in the AAAH model~\cite{vadim, altshuler} at {\it infinite temperature}. The infinite temperature limit is defined by averaging observables over all energy eigenstates (i.e. with a thermal weighting factor equal to unity). 
 For the non-interacting GAA model with a SPME, the average is performed over localized, extended, and partially-extended states, and thus leading to extended behavior, e.g., the averaged  EE  obeys volume law. 
 For the interacting case, 
 in contrast to
MBL studies in the absence of SPME where 
 interactions  
 make the system more extended, we 
 find that 
 interactions could actually stabilize the infinite temperature MBL phase in a considerable parameter region of the model in Eq.(\ref{eq:Ham}). 

\begin{figure}[htp!] 
\includegraphics[angle=0,width=.8\linewidth]{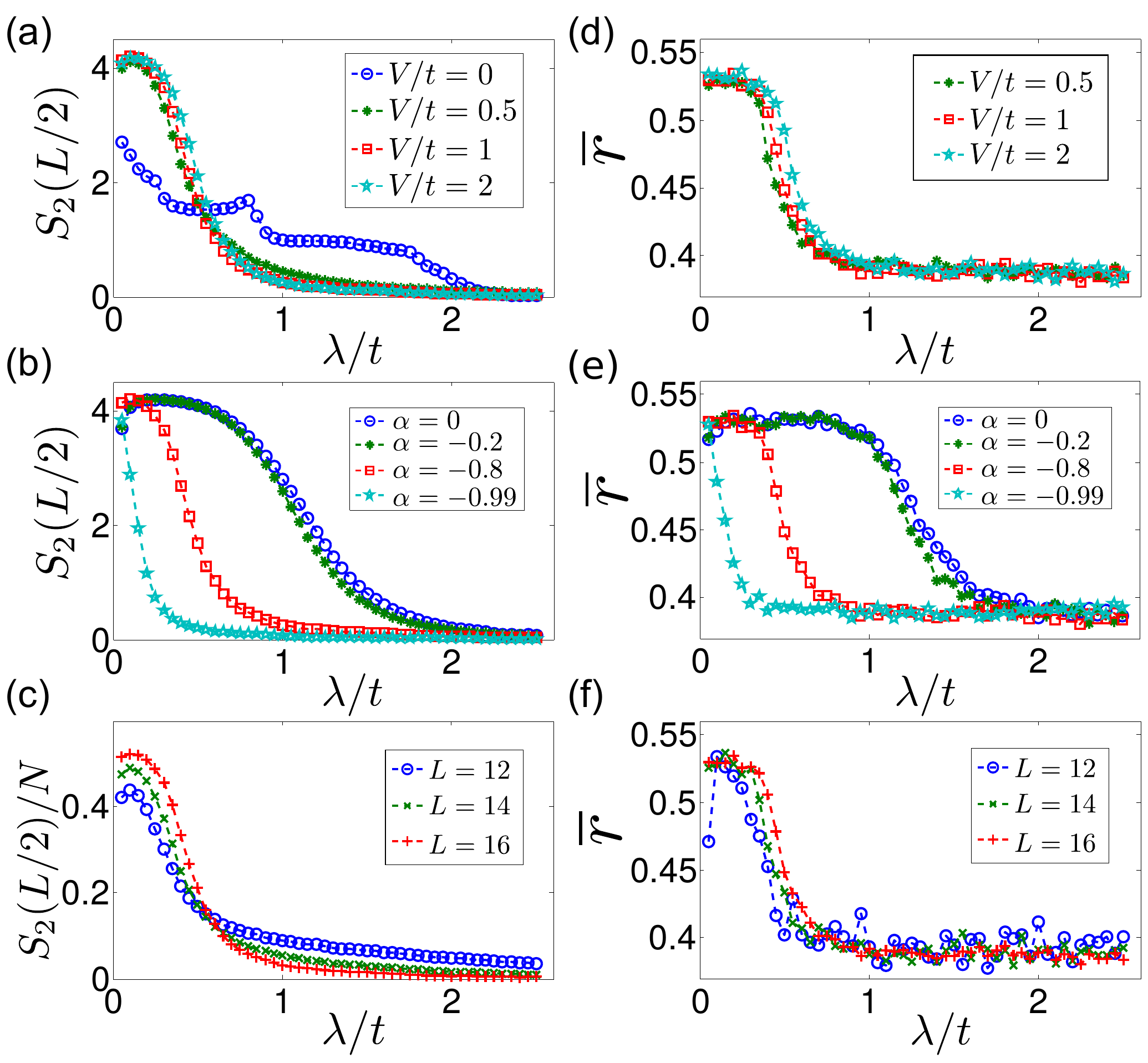}
 \caption{MBL phase at half filling and infinite temperature. (a), (b) and (c) show the R\'enyi entropy  versus the incommensurate potential strength. (d), (e) and (f) show the averaged adjacent gap ratio. In (a) and (d), we show the interaction dependence with $\alpha=-0.8$. The numerical results indicate the interacting system is in the MBL phase (all states localized) when $\lambda$ is larger than a certain critical value ($\lambda_c \approx 0.8$). In particular (a) explicitly shows interaction effects make the GAA model more localized. 
 (b) and (e) show
 the $\alpha$ dependence with $V/t = 1$. As we increase $\alpha$, $\lambda_c$ becomes smaller. In (c) and (f), we show the system size dependence with $V/t = 1$ and $\alpha = -0.8$. 
}
\label{fig:Fig3} 
\end{figure} 

To capture the MBL transition at infinite temperature, in addition to the  EE  scaling (averaged over all eigenstates), we present the level statistics, an established diagnostic for MBL~\cite{vadim,oganesyan2007}.
We consider the dimensionless adjacent gap ratio, 
\be
\textstyle r_n = \rm {min} (\delta_n, \delta_{n+1})/{\rm max} (\delta_n, \delta_{n+1}), \nn 
\ee 
where $\delta_n = E_{n+1}-E_n$. 
We calculate its average $\overline{r} = \frac{1}{V_H} \sum_n r_n$ 
($V_H$ is the Hilbert space dimension), to locate the MBL phase~\cite{oganesyan2007}.

As shown in Fig.~\ref{fig:Fig3}, 
when the incommensurate potential strength $\lambda$ is weak,  the  EE  is extensive, a signature for the system being extended. The average adjacent gap ratio is $\overline{r}\approx 0.53$,  
which implies that the energy spectra satisfy 
Gaussian orthogonal ensemble level statistics,
and the many-body phase is delocalized~\cite{1993_Shklovskii_PRB,oganesyan2007,vadim}. 
When $\lambda$ is above a certain threshold $\lambda_{c}(\alpha, V)$, the model undergoes the MBL transition. In this parameter regime, 
the  EE  obeys area law scaling (Fig.~\ref{fig:Fig3}c), and the gap ratio becomes $\overline{r}=0.39$ (Fig.~\ref{fig:Fig3}f), which is consistent with eigenstates satisfying a Poisson distribution, and the model is in the MBL phase~\cite{1993_Shklovskii_PRB,oganesyan2007,vadim,2013_Bauer_JSM}. 
We emphasize that for $\lambda > \lambda_c $ in the presence of delocalized single-particle orbitals, the MBL phase can still be stabilized by interactions. For example, with $\lambda/t = 1.5$ and $\alpha = -0.80$, although the non-interacting case is not completely localized due to the SPME, the interacting system is completely localized as implied by the  EE  (Fig.~\ref{fig:Fig3}a), adjacent gap ratio (Fig.~\ref{fig:Fig3}d), and their system size dependence (Fig.~\ref{fig:Fig3}(c,f)). In addition, 
starting from $\alpha=0$ (with no SPME) as $\alpha\rightarrow -0.99$, where more localized orbitals are mixed in, 
we find both $S_2(L/2)$ and $\bar{r}$  decrease as displayed in Fig.~\ref{fig:Fig3}(b,e). 
To conclude, we have established the existence of an MBL phase at infinite temperature in the presence of SPME.

{\it Experiment.} To study the MBL phase in the AAAH model, a two-component Fermi gas of $K^{40}$ atoms has been recently confined in a 1d superlattice with optical potential, 
$V_0 \cos^2 (k x) + V_1 \cos^2 (k' x )$~\cite{blochmbl}, with $k'$ incommensurate to $k$. 
To investigate our predicted mobility edge physics, we propose to add an additional potential, $V_2 \cos ^2 (2k' x)$. Choosing $V_0 = 5 E_r$, $V_1 = 0.13 E_r$, and $V_2 = 0.026 E_r$, ($E_r$ is the single-photon recoil energy)~\cite{2008_Bloch_RMP}, the non-interacting model in Eq.~\eqref{eq:Ham} with $\lambda/t  = 1$, and $\alpha = 0.2$ is approximately realized~\cite{citesupplement}, 
with localization properties described in 
Ref.~\cite{sriramgaa}. 
Preparing an initial state with its average energy in the non-thermal extended region, 
its unitary evolution 
would provide direct observation of nontrivial 
relaxation of an interacting many-body state in presence of SPME~\cite{2008_Rigol_thermalization_Nature,2011_Canovi_PRB}. 
As the numerical simulations on classical computers are limited in terms of system size, the quantum simulator, atoms in the optical lattice, would clarify the thermodynamic limit of our proposed non-ergodic and MBL phenomena.

 {\it Conclusions:-} 
In summary, we have shown the single-particle mobility edge and interactions result in a many-body 
mobility edge. A central new result here is the existence of two characteristic many-body energies (i.e. $E_L$ and $E_T$) in general in a system with a corresponding SPME, which separate localized and  extended  states ($E_L$) and non-ergodic and thermal states ($E_T$). 
Our numerical results (within our numerical accuracy and within the finite size limitations) suggest $E_L < E_T$, which allows  
for the possibility of non-ergodic delocalized many-body states (i.e. a non-ergodic metal) as a strange new intermediate phase of quantum matter. We expect our findings to generically apply to systems with SPME, specifically, to the three dimensional interacting disordered Anderson model.

\acknowledgments{
{\it Note added.}
After completing our work we became aware of a complementary and independent recent study~\cite{modak2015many} of many-body localization in systems with mobility edges.

{\it Acknowledgements.} This work is supported by JQI-NSF-PFC, ARO-Atomtronics-MURI and LPS-CMTC. We would like to thank Yang-Le Wu, Yi Zhang,  Arijeet Pal, Yang-Zhi Chou, and Matthew Foster for useful discussions. X.L. would like to thank the Department of Energy's Institute for Nuclear Theory at the University of Washington for its hospitality during the completion of this work. 
}

\bibliographystyle{my-refs}
\bibliography{references.bib}

\begin{thebibliography}{10}
\bibitem{citationnote}
This follows since the many-body states with energy above $E_B$ [below $E_A $]
  are all extended or partially-extended [localized] by definition. If $E_B <
  E_A$, no states are allowed between $E_B$ and $E_A$, and the
  (partially)-extended and localized states are thus separated by an energy gap
  $E_A - E_B $. Such an energy gap closes when $E_B = E_A $. When $E_B > E_A $,
  states are allowed between $E_A $ and $E_B $. We can reasonably assume for
  the model we consider that there is no single-particle energy gap near the
  single-particle mobility edge or at least, any such gap is nongeneric and
  negigibly small. Then the many-body energy spectrum is typically dense and
  does not involve a gap at finite energy density. It follows then that $E_B >
  E_A$ is most typical.
\end{thebibliography}

\onecolumngrid
\newpage 

\appendix

\renewcommand{\figurename}{{\bf Supplementary Figure}}

\begin{center} 
{\Large \bf Supplementary Material: \\
Many body localization and quantum non-ergodicity in a model with a single-particle mobility edge}  
\end{center} 

\section{Fraction of delocalized states as a function of ($\alpha$, $\lambda$).}

In this section we compute the fraction of localized states as a function of $\lambda$ for different values of $\alpha$ considered in the main text. In Supplementary Figure~\ref{ratio}, we plot the ratio of the delocalized states to the total number of states as a function of $\lambda/t$ for different values of $\alpha$= (0, -0.2, -0.8, -0.99). For $\alpha=0$, all the states are either delocalized or localized  as a function of $\lambda/t$ (with $\lambda/t=1.0$ being the critical point). For $\alpha\ne 0$, there is an energy dependence to the fraction of delocalized states and the many step structure is in part due to the gap in the single particle energy spectrum. Thus the parameter $\alpha$ allows us direct control of the fraction of single particle localized (delocalized) states participating the the non-interacting many body state given by the slater determinant.
 \begin{figure}[htb!]
  \centering
\includegraphics[width=6cm,height=5.25cm]{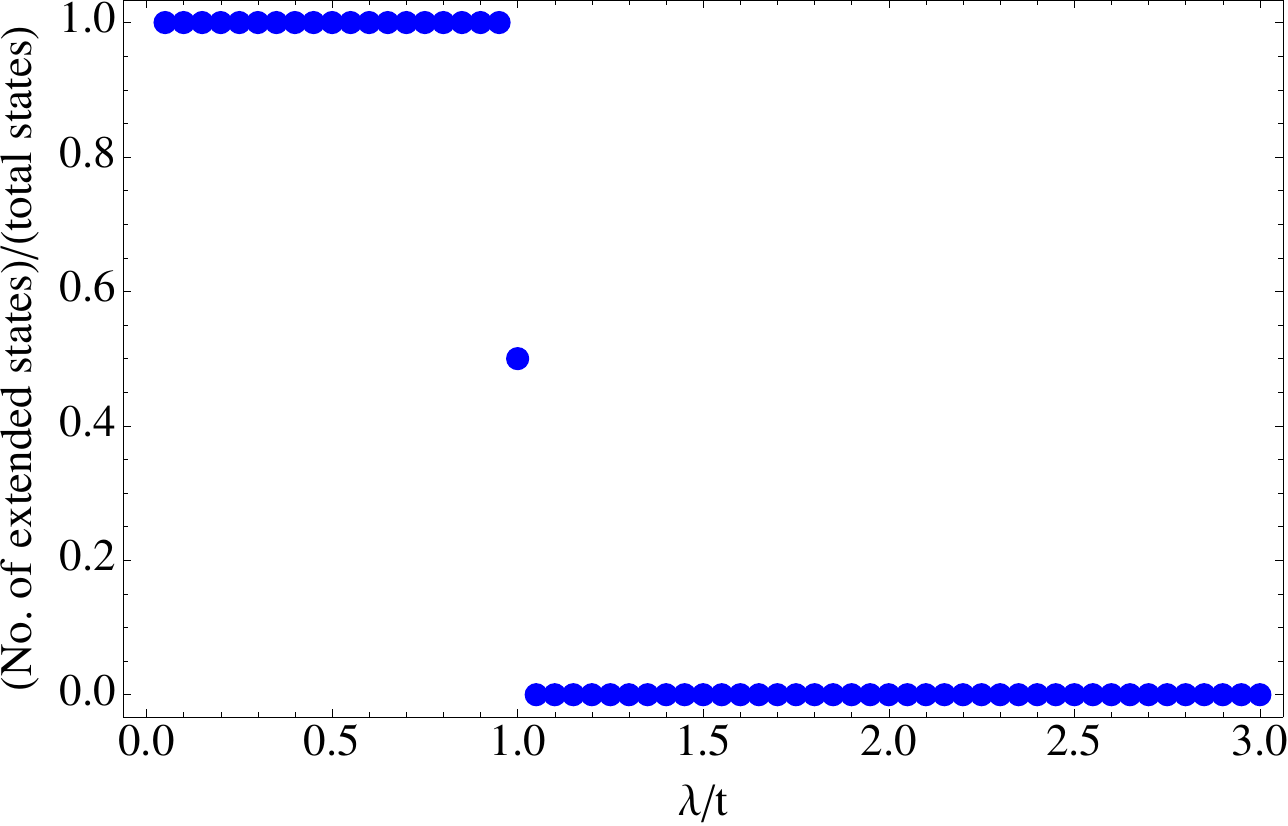}\hspace{0.1cm}\includegraphics[width=6.0cm,height=5.25cm]{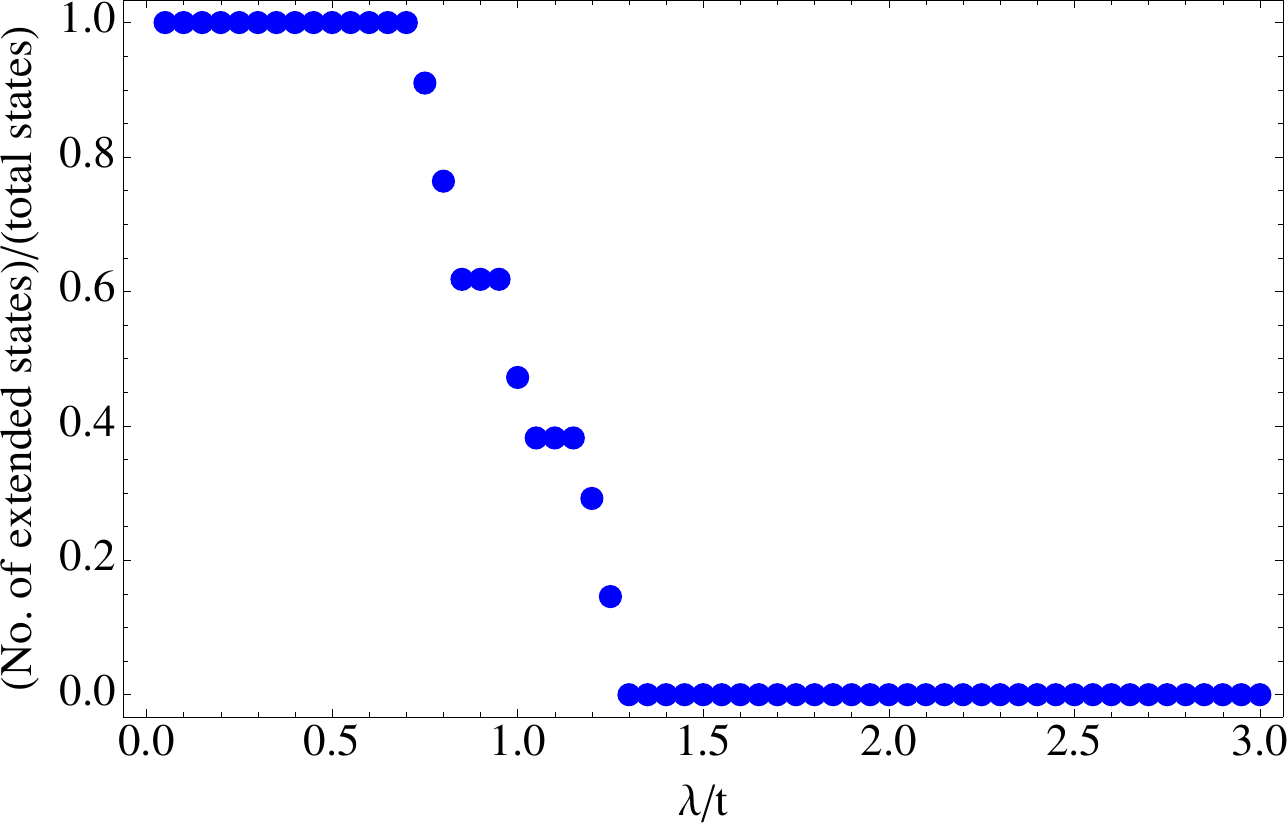}\\
\hspace{0.7cm}\textrm{(a)}\hspace{5.5cm}\textrm{(b)}\\
\includegraphics[width=6.0cm,height=5.25cm]{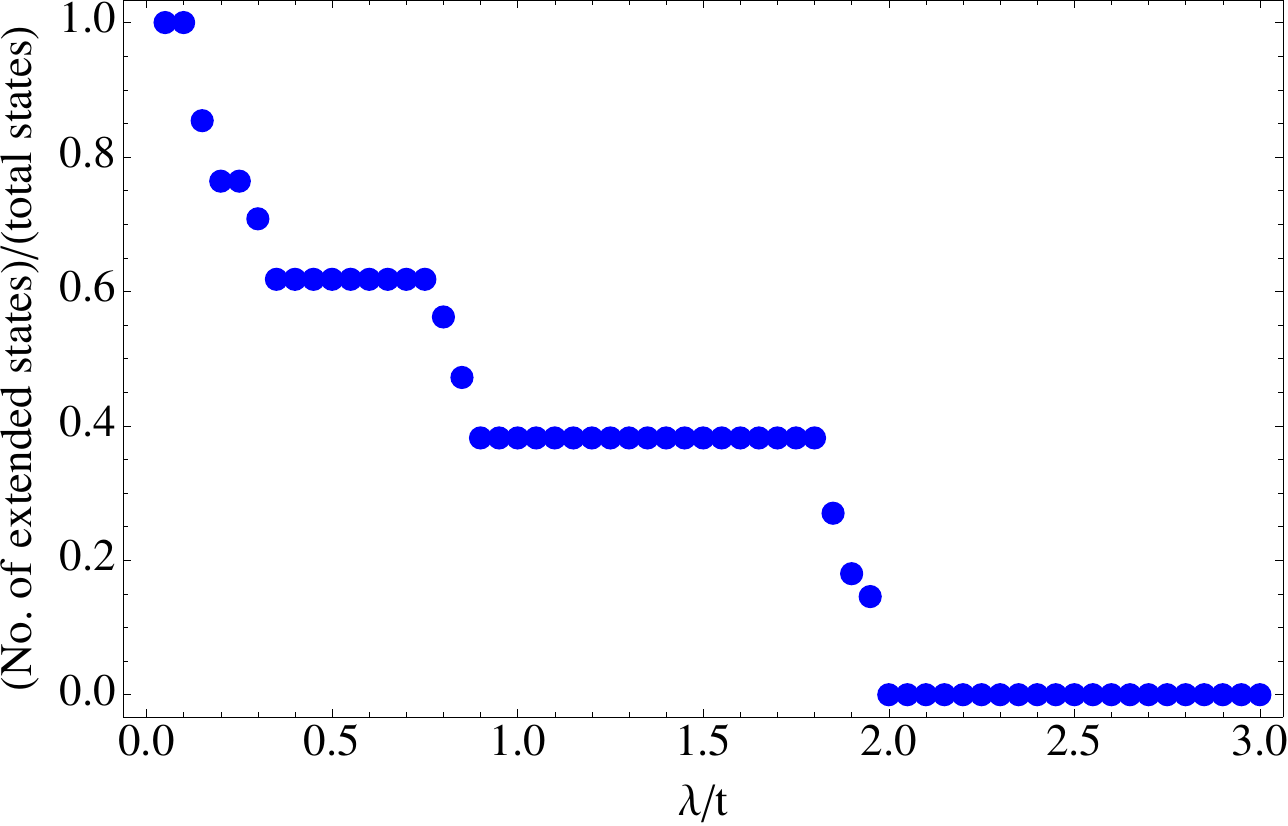}\hspace{0.1cm}\includegraphics[width=6.0cm,height=5.25cm]{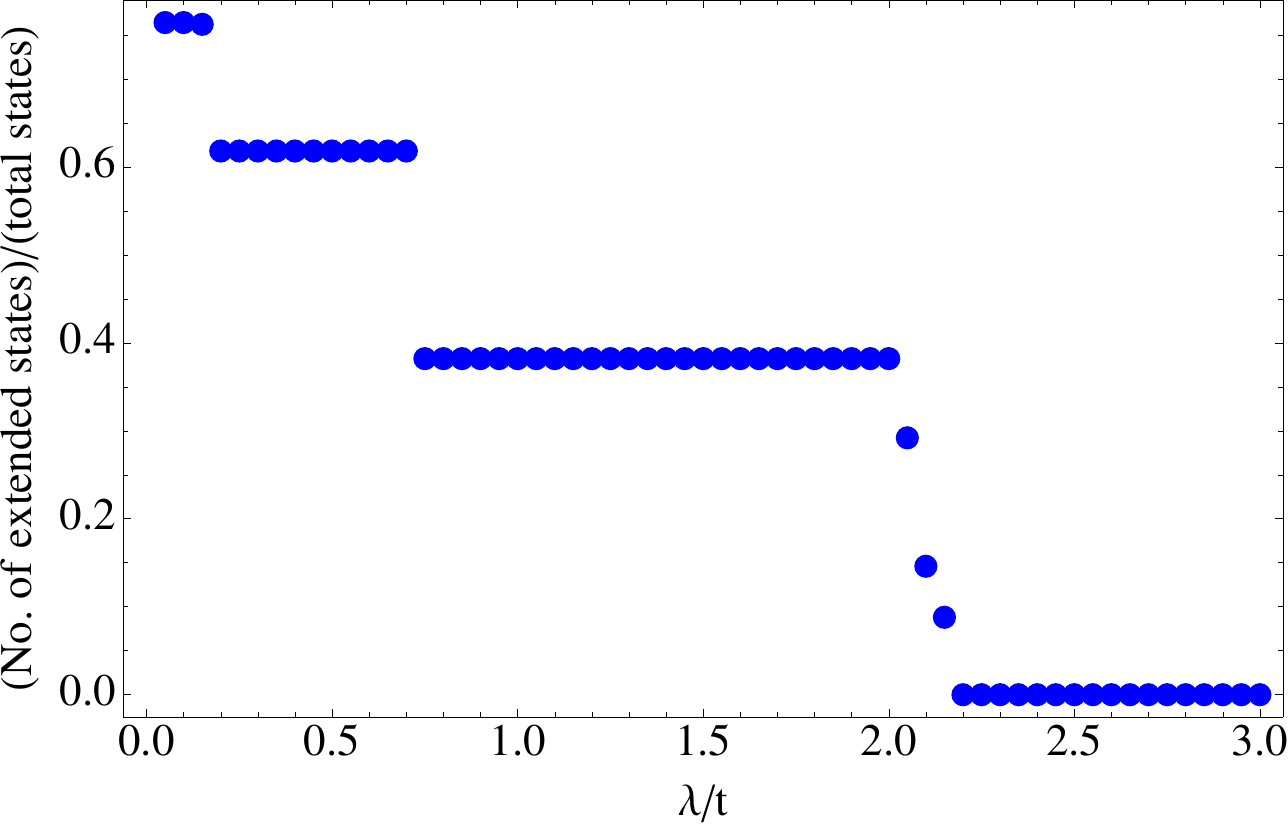}\\
\hspace{0.7cm}\textrm{(c)}\hspace{5.5cm}\textrm{(d)}
    \caption{Number of delocalized state to the total number of states as a function of $\lambda/t$ for different values of $\alpha$= (0, -0.2, -0.8, -0.99)}
  \la{ratio}
\end{figure}

\begin{figure}[htp] 
\includegraphics[angle=0,width=.8\linewidth]{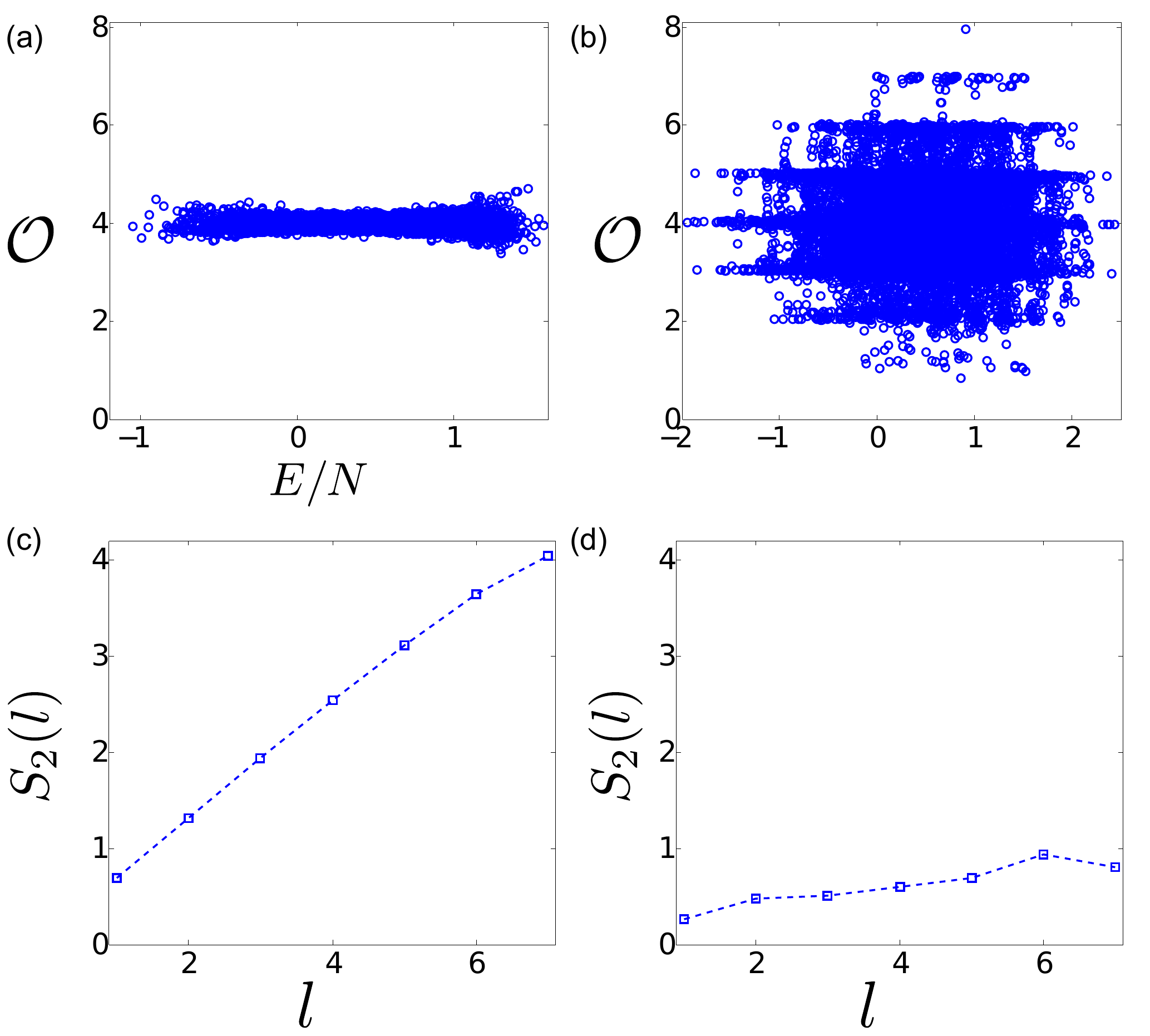}
\caption{
Completely localized and extended phases in the AAAH model. (a) and (b) show the fluctuation in the observable ${\cal O} (E)$ in the extended and localized phases, respectively. (c) and (d) show the corresponding R\'enyi entropy scaling. Here, in calculating entanglement entropy, we average over all eigenstates. In this plot we fix $V/t = 1$, $\alpha= 0$, $N = 8$, $L = 16$, and we use $\lambda/t = 0.3$ and  $\lambda/t = 1.5$, respectively for the extended and localized phases. In the localized phase (b,d), we have large fluctuation in ${\cal O} (E)$ for nearby eigenstates, a signature for non-ergodicity, and the entanglement entropy obeys area law. In the extended phase (a,c), the fluctuation in ${\cal O} (E)$ is suppressed, and the entanglement entropy obeys volume law.   
}
\label{fig:OS2AA} 
\end{figure}

\section{Details of localized states with interactions} 
In Supplementary Figure~\ref{fig:OS2AA}, we provide more results about the observable ${\cal O} (E) = \sum_{j=1}^{L/2}  n_j$ showing its behavior in completely localized and extended phases, where the non-ergodicity is consistent with the entanglement entropy scaling. 

In numerics, the fluctuation of observables (${\cal O}(E)$) between nearby eigenstates, 
is quantified by the standard deviation of ${\cal O}(E)$, ${\rm var}[{\cal O}]$, over some small finite energy range 
$[E-\delta E/2, E+\delta E/2]$ (we choose $\delta E/(Nt) = 0.02$ in this work). 

In averaging over the phase $\phi$ ({\it see Eq.(1) in the main text}), we choose $10$ realizations with $\phi/\pi = \{0, 0.2, \ldots 1.8\}$. We first calculate ${\rm var} [{\cal O}]$ for each $\phi$-realization and then take the average. As the energy spectra for different $\phi$ are not identical, we actually average over different data points within an energy window set by $\delta E'/(Nt) = 0.04$.

\section{Detailed description about experimental realization} 
In experiments to probe the MBL phase in the AAAH quasiperiodic model, a 1d superlattice is engineered with counterpropagating laser beams, and the resultant optical potential reads 
$V_0 \cos^2 (k x) + V_1 \cos^2 (k' x )$~\cite{blochmbl}, with $k'$ incommensurate to $k$. 
With $V_0\gg E_r$ ($E_r$ is the single photon recoil energy), the tight binding model is a valid description for the experimental system. Further requiring $0<|V_1| \ll V_0$, the AAAH model is reached. Taking $V_0  = 5 E_r$, the parameters in the AAAH model are then given by  
($t \approx 0.065E_r$, $\lambda/t \approx \frac{1}{2} V_1/t$). 

Considering the limit  $|\alpha|\ll 1$ of the GAA model (see {\it Eq.~(1) of the main text}), the on-site potential becomes  $\sim 2\lambda(\cos (2\pi j k+\phi)+ \alpha \cos^2(2\pi j k+\phi))$. It follows that  to implement the GAA model we need an additional potential, $V_2 \cos ^2 (2k' x)$. Using $V_2 \ll V_1$, the GAA model is approximately realized with $\alpha$ estimated to be $\alpha \approx V_2/V_1$. Choosing $\lambda = t$, the small $\alpha$ limit supports both of sharply localized and delocalized wavefunctions (see Ref.~\cite{sriramgaa}), and there is a well-defined single-particle mobility edge. It is worth noting here that the $\alpha \ll 1$ limit may or may not be an experimental sweet spot to probe the mobility edge physics, as it  requires fine tuning of the laser strength to get $\lambda = t$. Searching for the parameter region that is most suitable for experiments is not the focus of this work and is thus left for future.

\section{Normalized participation ratio for non-interacting many body states} 
To describe a many body state, we can choose a Fock state basis 
$$ 
|j_1 j_2 \ldots j_N  \rangle = c_{j_1} ^\dag c_{j_2} ^\dag \ldots c_{j_N} ^\dag |{\rm vac}\rangle, 
$$ 
with $N$ the particle number. We will take an ordering $j_1<j_2 < \ldots j_N$. 
The many body wavefunction in this basis is 
\be 
| \Psi\rangle = \sum_{j_1< j_2< \ldots< j_N  } 
  \Psi_{\{j_1, j_2, \ldots, j_N \}} 
    | j_1, j_2, \ldots j_N   \rangle, 
\ee 
from which the NPR  is defined to be~\cite{vadim}
\be 
\eta (|\Psi\rangle)    = \frac{1}{\sum_{j_1< j_2< \ldots <j_N  }  |\Psi_{\{j_1, j_2, \ldots j_N\} }  | ^4 V_H }, 
\ee 
with $V_H$ the dimension of the many body Hilbert space,  which is 
$\left( \begin{array}{c}
L \\ 
N
\end{array}
\right)$ .

Without interaction, the many body eigenstate is simply a product state of single-particle orbitals, namely 
\be 
|\Psi\rangle ^ {\rm free} = |m_1, m_2, \ldots, m_N \rangle 
= \psi_{m_1} ^\dag \psi_{m_2} ^\dag \ldots \psi_{m_N} ^\dag |{\rm vac}\rangle.  
\label{eq:psistate}
\ee 
First, as constructed in Eq.~\eqref{eq:psistate}, the many body wavefunction is given as 
\be 
\Psi_{\{j_1, j_2, \ldots j_N\}}  ^{\rm free} 
 = \sum_P {\rm sgn}(P) \psi_{m_1 } ( j_{P[1]})  \ldots \psi_{m_N} (j_{P[N]} ), 
\ee 
where $P$ labels all permutations. 
Now we get 
\bea 
I_4 (|m_1, m_2, \ldots, m_4 \rangle)   &\equiv & \sum_{ j_1< j_2< \ldots <j_N } 
| \Psi_{ \{j_1, j_2, \ldots j_N\} } ^{\rm free} |^4 \nn  \\ 
 &=& \sum_{P_a, P_b, P_c, P_d} {\rm sgn} (P_a P_b P_c P_d) 
 \sum_{ j_1< j_2< \ldots <j_N }  
 \left[ \psi_{m_1} ^* (j_{P_a[1]} ) \ldots \psi_{m_N} ^* (j_{P_a[N]})  \right] \nn \\
&& \times  \left[ \psi_{m_1} ^* (j_{P_b[1]} ) \ldots \psi_{m_N} ^* (j_{P_b[N]}) \right] 
 \left[ \psi_{m_1} (j_{P_c[1]} ) \ldots \psi_{m_N} (j_{P_c[N]}) \right] 
 \left[ \psi_{m_1} (j_{P_d[1]} ) \ldots \psi_{m_N} (j_{P_d[N]}) \right] \nn \\ 
 & = & 
 \sum_{P_a', P_b', P_c'} {\rm sgn} (P_a' P_b' P_c')   \sum_{ j_1, j_2, \ldots, j_N } 
 \left[ \psi_{m_1} ^* (j_{P_a'[1]} ) \ldots \psi_{m_N} ^* (j_{P_a'[N]})  \right] 
 \left[ \psi_{m_1} ^* (j_{P_b'[1]} ) \ldots \psi_{m_N} ^* (j_{P_b'[N]}) \right] \nn \\
 && \times \left[ \psi_{m_1} (j_{P_c'[1]} ) \ldots \psi_{m_N} (j_{P_c'[N]}) \right] 
 \left[ \psi_{m_1} (j_1 ) \ldots \psi_{m_N} (j_{N}) \right].
\eea

If one orbital, say $m_1$, is localized, we take $\psi_{m_1} (j) \sim \delta _{j, 1} $ (up to a local unitary transformation), 
and we  get 
\bea
I_4   &=& \sum_{\tilde{P}_a, \tilde{P}_b, \tilde{P}_c} 
{\rm sgn} (\tilde{P}_a \tilde{P}_b  \tilde{P}_c) 
\sum_{ j_2, j_3 \ldots, j_N } 
 \left[ \psi_{m_2} ^* (j_{\tilde{P}_a [2]} ) \ldots \psi_{m_N} ^* (j_{\tilde{P}_a [N]})  \right] 
  \left[ \psi_{m_2} ^* (j_{\tilde{P}_b [2]} ) \ldots \psi_{m_N} ^* (j_{\tilde{P}_b [N]}) \right]  \nn \\ 
&& \times \left[ \psi_{m_2} (j_{\tilde{P}_c [2]} ) \ldots \psi_{m_N} (j_{\tilde{P}_c [N]}) \right] 
 \left[ \psi_{m_2} (j_2 ) \ldots \psi_{m_N} (j_{N}) \right], 
\eea 
where $\tilde{P}$ means a permutation among the numbers $(2, 3, 4, \ldots, N)$. It follows that  we have  
a particle number reduction relation 
\be 
I_4  (|m_1, m_2 , \ldots, m_N \rangle )= I_4 ( |m_2, m_3 , \ldots, m_N \rangle ). 
\label{eq:reduction} 
\ee 
The Hilbert space dimension for $N-1$ fermions on $L$ sites is 
$\left( 
\begin{array}{c}
L \\ 
N-1
\end{array}
\right)$. 
We thus know that at low filling 
$$ 
I_4 (|m_2, m_3, \ldots, m_N \rangle ) \ge 
\left( 
\begin{array}{c}
L \\ 
N-1
\end{array}
\right)^{-1} 
\gg 
\left( \begin{array}{c}
L \\ 
N
\end{array}
\right)^{-1} .
$$ 
Then we have $ 
\eta ( |m_1, m_2, \ldots, m_N \rangle) \to 0, 
$ as $L \to \infty$, if we fix the particle number.

Suppose we have a number, $N_{\rm loc}$, of single particle localized orbitals, $m_1$, $m_2$, \ldots, and $m_{N_{\rm loc}} $, then it follows from the reduction relation (Eq.~\eqref{eq:reduction}) that 
\be 
I_4( |m_1, m_2, \ldots, m_N \rangle ) = I_4 ( | m_{N_{\rm loc}+1}, \ldots m_N \rangle ). 
\label{eq:particlereduction} 
\ee 
Then the NPR value satisfies 
\be 
\eta (|m_1, m_2, \ldots, m_N \rangle) \le \left( \frac{N}{L-N} \right) ^{N_{\rm loc}} . 
\ee 
Considering the thermodynamic limit, $L \to \infty$ while keeping $N/L = \rho_{\rm tot}$, 
$ \eta \le \left(\frac{\rho_{\rm tot}}{1-\rho_{\rm tot} } \right)^{N_{\rm loc}}$.  To have $\eta \to 0$, 
the number of localized single-particle orbitals composing the many body states is required to be also extensive, i.e., 
$N_{\rm loc} = \rho_{\rm loc} L$ (in principle, $\rho_{\rm loc}$ could be an infinitesimal number), 
as we take the thermodynamic limit. 
Our conclusion is, if the many body state is composed of some finite (extensive) fraction 
of localized single-particle orbitals, then its NPR value vanishes. The reverse is also true at low filling. 
The most general case shall be discussed elsewhere.

The above analysis based on NPR shows that a many body state composed of both localized and extended orbitals behaves like a localized state, which is opposite to using the entanglement entropy scaling criterion.  Nonetheless, when the system is in a MBL phase with all states localized, NPR and entanglement entropy give consistent results (see Supplementary Figure~\ref{fig:NPRresults} and {\it Figure $3$ in the main text}). For example, with   $\lambda/t = 1.5$, $\alpha = -0.8$, and $V/t = 1$, the three different diagnostics, including NPR, entanglement entropy and adjacent gap ratio consistently support all many body states are localized.

\begin{figure}[htp] 
\includegraphics[angle=0,width=\linewidth]{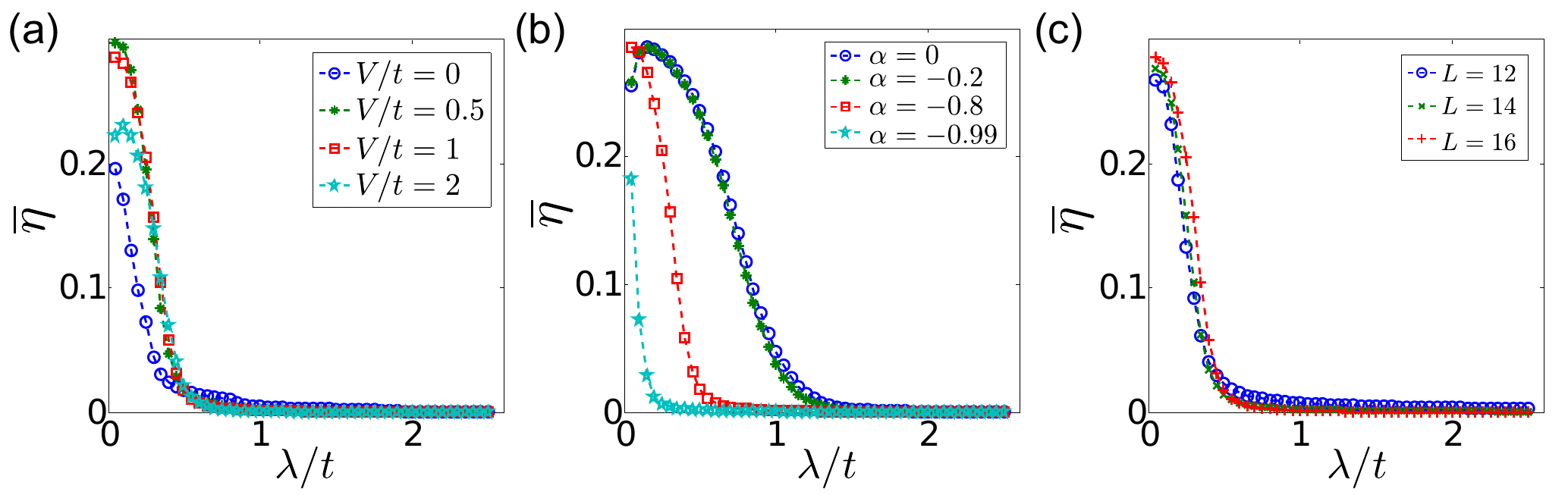}
\caption{The normalized participation ratio as a function of $\lambda$ for the GAA model. (a) shows the interaction dependence with $\alpha = -0.8$. (b) shows the $\alpha$ dependence with $V/t = 1$. (c) shows the finite size dependence with $V/t = 1$ and $\alpha = -0.8$.  We choose $\phi = 0$ and $N/L = 1/2$ in this plot.  
}
\label{fig:NPRresults} 
\end{figure}


\section{Similarity of the GAA to 3d Anderson model} 
In Supplementary Figure~\ref{fig:GAAvsAnderson}, we show the Von Neumann entropy for single-particle states in 1d GAA and 3d Anderson models. The behavior of localized and extended states appears similar in these two models. We suspect  that the many body localization phenomena in the 1d GAA model could resemble that in the 3d Anderson model. Detailed comparison of these two models will be discussed elsewhere (work in progress).

\begin{figure}[htp] 
\includegraphics[angle=0,width=\linewidth]{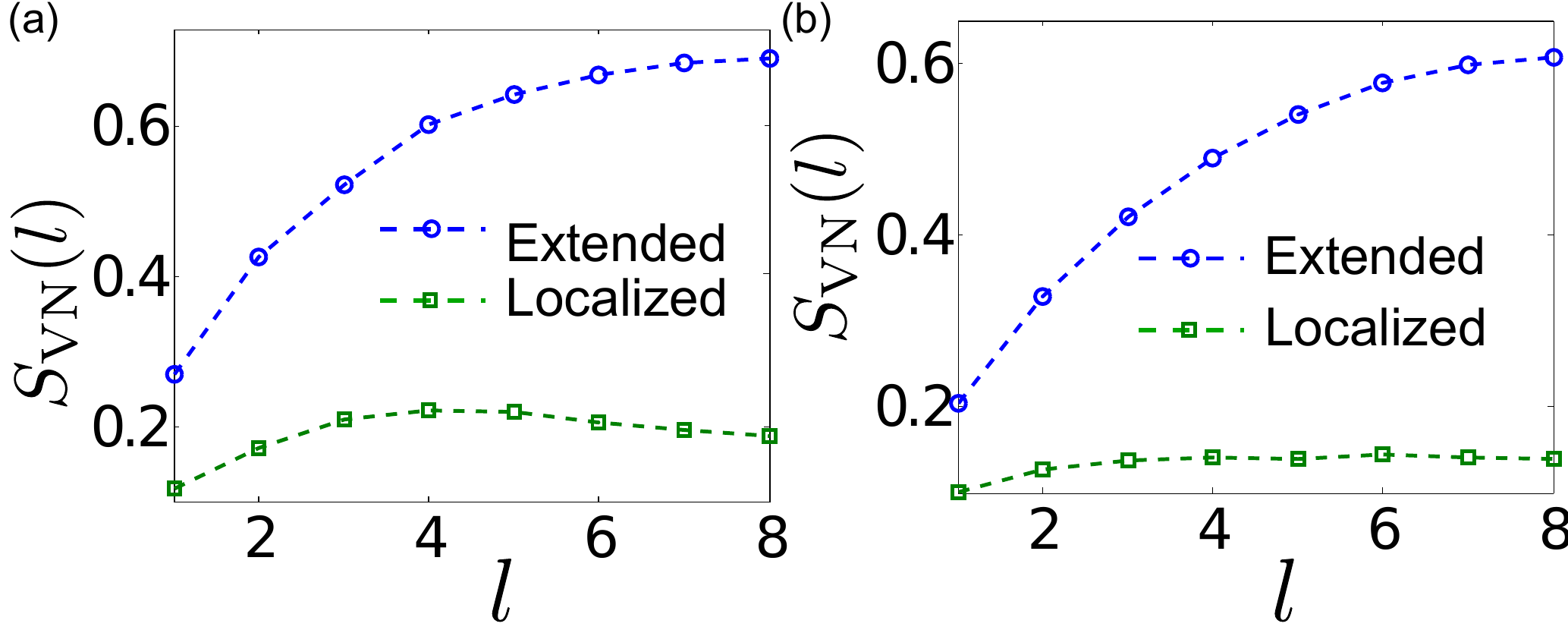}
\caption{
Von Neumann entropy for single-particle extended and localized states. (a) and (b) show the results obtained from the 1d GAA and 3d Anderson models~\cite{AndersonLocalization}, respectively. For the GAA model, we choose $\lambda/t = 0.3$ and $\alpha = -0.8$, and the phase $\phi$ is averaged over. For the 3d Anderson model, we choose the disorder strength to be sixteen times of the nearest neighbor tunneling, and we average over  two thousand disorder realizations. For both localized and extended states, we simply pick certain energies, and the qualitative behavior of the entanglement entropy does not depend on the details. 
}
\label{fig:GAAvsAnderson} 
\end{figure}

\end{document}